\documentclass[conference]{IEEEtran}
\ifCLASSINFOpdf
\else
\fi
\usepackage{ntheorem,amsmath,graphicx,float,caption}
\usepackage{subeqnarray,cases,flushend,fancyhdr,subfigure}
\newtheorem{Example}{Example}
\newtheorem{Simulation}{Simulation}
\begin{document}
%
\title{The Rationale for Second Level Adaptation}

\pagestyle{fancy}
\author{Kumpati S. Narendra, Yu Wang and Wei Chen\\
\\
{\large Center for Systems Science, Yale University}\\
}


%


\maketitle

\begin{abstract}
Recently, a new approach to the adaptive control of linear time-invariant plants with unknown parameters (referred to as second level adaptation), was introduced by Han and Narendra in [1]. Based on $N (\geq m+1)$ fixed or adaptive models of the plant, where $m$ is the dimension of the unknown parameter vector, an unknown parameter vector $\alpha\in R^{N}$ is estimated in the new approach, and in turn, is used to control the overall system. Simulation studies were presented in [1] to demonstrate that the new method is significantly better than those that are currently in use.

In this paper, we undertake a more detailed examination of the theoretical and practical advantages claimed for the new method. In particular, the need for many models, the proof of stability, and the improvement in performance and robustness are explored in depth both theoretically and experimentally.
\end{abstract}


%
\IEEEpeerreviewmaketitle

\section{Introduction}
For over two decades, there has been a great deal of interest in developing new methods for rapidly identifying and controlling linear time-invariant plants with large parametric uncertainties in a stable fashion. It is argued that such methods can also be used to adaptively control linear systems with time-varying parameters without losing stability. Recently, second level adaptation was proposed in [1], where the information generated by multiple adaptive models can be used at a second level to identify the plant rapidly. If the unknown plant parameter vector is of dimension $m$, and belongs to a compact set, it is well known that $(m+1)$ points can be chosen in parameter space so that the unknown vector lies in their convex hull. This is the starting point of second level adaptation. If $m+1$ models are chosen with these parameters as their initial values, it was argued in [1] that the information derived from them could be used to derive an equivalent parameterization of the plant which converges much more rapidly. The basic ideas involved in the new  approach can be summarized as follows, for use throughout the paper:
\begin{enumerate}
\item The set $S_{\theta}\subset R^{m}$ in parameter space in which the unknown parameter $\theta_{p}$ can lie, is convex.
\item Parameters $\theta_{1}(t_{0}), \theta_{2}(t_{0}),\cdots, \theta_{m+1}(t_{0})$ can be chosen such that $\theta_{p}$  lies in the convex hull of $\theta_{i}(t_{0}),\{i=1,2,\cdots,m+1\}$. This implies that constants $\alpha_{1},\alpha_{2},\cdots,\alpha_{m+1}$ with $0\leq \alpha_{i}\leq 1$ and $\sum_{i=1}^{m+1}\alpha_{i}=1$ exits such that $\sum_{i=1}^{m+1}\alpha_{i}\theta_{i}(t_{0})=\theta_{p}, \theta_{p}\in\Omega$.
\item If $(m+1)$ adaptive models are used to identify $\theta_{p}$ at every instant, it can be shown using arguments of linearity and convexity that $\theta_{p}$ also lies in the convex  hull of $\theta_{i}(t)$, ($i.e. \sum_{i=1}^{m+1}\alpha_{i}\theta_{i}(t)=\theta_{p}$ and $\theta_{i}(t) (i=1,2,\cdots,n+1)$ are the estimates of $\theta_{p}$ generated by the adaptive models) provided the initial states of the plant and models are identical. It is worth noting that when the state variables of the plant are accessible, the initial conditions of the states of the models can be chosen to be identical to that of the plant.
\item If the $(m+1)$ models are non-adaptive but constant (with $\theta_{i}$ constant), $ \sum_{i=1}^{m+1}\alpha_{i}e_{i}(t)\equiv0$, where $e_{i}(t)$ are the $(m+1)$ identification errors.
\item Since $\theta_{i}(t)$ in (\ref{hypercubic}) and $e_{i}(t)$ in (\ref{basic reference model}) are accessible, the conditions
\begin{equation}
\begin{split}
\sum_{i=1}^{m+1}\alpha_{i}\theta_{i}(t)=&\theta_{p}\\
\text{or}\ \ \ \sum_{i=1}^{m+1}\alpha_{i}e_{i}(t)=&0
\end{split}
\end{equation}
can be used to estimate the values of the constants $\alpha_{i}$. Since by equation (1), the vector $\alpha$ can be considered as an alternative parameterization of the unknown plant, it can be used to estimate $\theta_{p}$ and hence the control input $u(\cdot)$.
\item The estimate $\hat{\alpha}(t)$ of $\alpha$ can be determined either algebraically, or by using a differential equation as described later.
\end{enumerate}

In this paper we will be interested in the following two principal questions:
\begin{enumerate}
\item[(\expandafter{\romannumeral1})] under what conditions would we expect second level adaptation to result in better performance than first level (or conventional) adaptation?
\item[(\expandafter{\romannumeral2})] why is the overall system stable even when control is based on the estimate $\hat{\alpha}(t)$, rather than the true value of $\alpha$  of the unknown parameter vector? ($i.e$ proof of stability)
\end{enumerate}

In addition, we will also be interested in questions related to design and performance such as:
\begin{enumerate}
\item[(\expandafter{\romannumeral3})] can the performance and robustness of second level adaptation be improved by increasing the number of models? If additional models are to be used, where should they be located for maximal marginal benefit?
\item[(\expandafter{\romannumeral4})] are fixed models or adaptive models to be preferred? Can rational reasons be provided for combining such models?
\end{enumerate}

Starting with a brief review of the theory of second level adaptation based on both adaptive and fixed models, we shall discuss the above questions in detail and  compare both theoretically and using simulations, the new approach with conventional first level adaptive control. The paper will conclude with simulations studies of adaptation in rapidly time-varying environments and a discussion of future directions for research.

\section{Mathematical Preliminaries}
The principal ideas involved in second level adaptation using multiple adaptive models and fixed models are discussed briefly in this section. These are used in all the discussions throughout the paper.

\subsection{Second Level Adaptation (Adaptive models)}
We first consider the simplest case where the plant to be controlled adaptively is in companion form, and all the state variables of the plant are accessible. The plant is described by the state equations:
\begin{equation}
\label{basic system model}
\Sigma_{p}:\ \dot{x}_{p}=A_{p}x_{p}(t)+bu(t)
\end{equation}
where $x_{p}(t)\in r^{n},u(t)\in R$ and $A_{p}$ and $b$ are in companion form and defined as
\begin{equation}
\label{canonical matrix}
A_{p}=\left[\begin{array}{lllll}0 & 1 & 0 & \cdots & 0\\0 & 0 & 1 & \cdots & 0\\ \vdots & \vdots & \vdots & \vdots & \vdots\\ 0 & 0 & 0 & \cdots & 1\\a_{1} & a_{2} & a_{3} & \cdots & a_{n}\end{array}\right], b=\left[\begin{array}{l}0\\0\\0\\\vdots \\1\end{array}\right]
\end{equation}

The parameters $\{a_{i}\}$ are unknown and it is assumed that the plant is unstable. If $\theta_{p}^{T}=[a_{1},a_{2},\cdots,a_{n}]$, we shall refer to $\theta_{p}$ as the unknown plant parameter vector. We further assume that $\theta_{p}$ belongs to a convex set $S_{\theta}$, where $S_{\theta}$ is a hypercube in parameter space defined by
\begin{equation}
\label{hypercubic}
S_{\theta}=\{\theta_{p}|\underline{\theta}_{p}\leq \theta_{p}\leq \bar{\theta}_{p}\}
\end{equation}

The reference model $\Sigma_{m}$ is also in companion form and defined by the differential equation
\begin{equation}
\label{basic reference model}
\Sigma_{m}:\ \dot{x}_{m}=A_{m}x_{m}(t)+br
\end{equation}
As in equation (\ref{canonical matrix}), the last row of $A_{m}$ is $\theta_{m}^{T}$ and is chosen by the designer to make $A_{m}$ stable. The reference input $r(t)$ is uniformly bounded and specified.

\subsection{Statement of the Problem}
The problem of adaptive control is to choose the input $u(t)$ in the presence of parameter uncertainty so that the output $x_{p}(t)$ of the plant follows asymptotically the output $x_{m}(t)$ of the reference model ($i.e. \lim_{t\to \infty}\|x_{p}(t)-x_{m}(t)\|=0$). Direct and indirect control are two distinct methods for achieving the above objective.

{\large \underline{Direct Control}}: In this case, the input $u(t)$ is chosen as $u(t)=k^{T}(t)x_{p}(t)+r(t)$, where $k(t)$ is adjusted adaptively. Since it is known a prior that a constant gain vector $k^{*}$ exists such that $\theta_{p}+k^{*}=\theta_{m}$, $k(t)$ has to be adjusted such that $\lim_{t\to\infty}k(t)=k^{*}$. If the equations describing the plant and the reference model are respectively
\begin{equation*}
\begin{split}
\dot{x}_{p}&=A_{p}x_{p}+bu\\
\dot{x}_{m}&=A_{m}x_{m}+br\\
A_{m}&=A_{p}+bk^{*T}
\end{split}
\end{equation*}
the input $u(t)$ is chosen as
\begin{equation*}
u(t)=r(t)+k^{T}(t)x_{p}(t)
\end{equation*}
where $\dot{\tilde{k}}=-e_{c}^{T}Pbx_{p}(t), \tilde{k}=k-k^{*}, A_{m}^{T}P+PA_{m}=-Q<0$. $e_{c}(t)=x_{p}(t)-x_{m}(t)$ is the control error. If $V(e_{c},\tilde{k})=e_{c}^{T}(t)Pe_{c}(t)+\tilde{k}^{T}\tilde{k}$ is a Lyapunov function candidate, it follows that
\begin{equation}
\frac{dV(t)}{dt}=-e_{c}^{T}Qe_{c}\leq 0
\end{equation}
Hence, the system is stable and $e_{c}(t)$ and $k(t)$ are bounded. Further, since $\dot{e}_{c}(t)$ is bounded, it follows that $\lim_{t\to\infty}e_{c}(t)=0$ and $\lim_{t\to\infty}\dot{\tilde{k}}(t)=0$ if the input $r(t)$ is persistently exciting.

\underline{Comment}: In direct control, the feedback parameter $k(t)$ is directly adjusted using the control error and system identification (or parameter estimation) is not involved.

{\large \underline{Indirect Control}}: In contrast to direct control described earlier, in indirect control, the unknown parameter vector $\theta_{p}$ is estimated at every instant as $\theta_{1}(t)$ and in turn used to control the system. To assure the stability of the estimator, the following model is used:
\begin{equation}
\label{basic identification model}
\dot{\hat{x}}_{p}(t)=A_{m}\hat{x}_{p}(t)+[A_{1}(t)-A_{m}]x_{p}(t)+bu
\end{equation}
where $A_{m}$ is a stable matrix, $A_{m}$ and $A_{1}$ (like $A_{p}$) are in companion form and the last rows of the two matrices are respectively $\theta_{m}^{T}$ and $\theta_{1}^{T}$. The later is an estimate of the unknown plant parameter vector $\theta_{p}$. At every instant, $\theta_{1}(t)$ is used to determine the control parameter vector $k(t)$ as
\begin{equation*}
\begin{split}
k(t)&=\theta_{m}-\theta_{1}(t)\\
&=\theta_{m}-[\theta_{p}+\phi_{1}(t)]\\
&=k^{*}-\phi_{1}(t)
\end{split}
\end{equation*}
where $\phi_{1}(t)$ is the error in the estimate of $\theta_{p}$.\\

Note that in equation (\ref{basic identification model}), $A_{m}$ is chosen to be the same as the matrix in the reference model (\ref{basic reference model}). However, it can be any stable matrix. It is chosen as in equation (\ref{basic reference model}) to make the control problem simple.

Defining $e_{1}(t)=\hat{x}_{p}(t)-x_{p}(t)$, we have
\begin{equation}
\label{identification error equation}
\dot{e}_{1}=A_{m}e_{1}(t)+b\phi_{1}^{T}(t)x_{p}(t)
\end{equation}
Using standard arguments of adaptive control, the adaptive law
\begin{equation}
\label{basic adaptive law}
\dot{\theta}_{1}(t)=\dot{\phi}_{1}(t)=-e_{1}^{T}Pbx_{p}(t)
\end{equation}
follows readily. The adaptive law is seen to be very similar to that derived from direct control, except that identification rather than control error is used in (\ref{basic adaptive law}).

Note that the control error is not used in the adjustment of the control parameter $k(t)$. The parametric erorr  $\phi_{1}(t)$ tends to zero when the input $r(t)$ is persistently exciting, and converges to a constant value (if $\phi_{1}^{T}x_{p}(t)$ tends to zero) when it is not. In both cases, the control error tends to zero as seen from equation (\ref{identification error equation}).

{\large \underline{Multiple Models (Adaptive)}}: We now consider $N$ models used simultaneously to estimate the unknown plant parameter vectors $\theta_{p}$. All of them have identical structures
\begin{equation*}
\dot{x}_{i}=A_{m}x_{i}+[A_{i}(t)-A_{m}]x_{p}(t)+bu
\end{equation*}
with $x_{i}(t_{0})=x_{p}(t_{0})$, $A_{i}(t)$ in companion form, and the last row of $A_{i}(t)$ equal to $\theta_{i}^{T}(t)$. This implies that the $N$ parameter error vectors $\phi_{i}(t)$ and identification error vectors $e_{i}(t)\ (i=1,2,\cdots,N)$ are generated at every instant with the equations
\begin{subequations}
\begin{numcases}
\cdot \dot{e}_{i}(t)=A_{m}e_{i}(t)+b\phi_{i}^{T}(t)x_{p}(t),\ \ \ e_{i}(t_{0})=0\\
\dot{\theta}_{i}(t)=\dot{\phi}_{i}(t)=-e_{i}^{T}(t)Pbx_{p}(t),\ \ \ \theta_{i}(t_{0})=\theta_{i0}
\end{numcases}
\end{subequations}
Since $\sum_{i=1}^{N}\alpha_{i}\theta_{i}(t_{0})=\theta_{p}$ or equivalently $\sum_{i=1}^{N}\alpha_{i}\phi_{i}(t_{0})=0$ from equation (10a), it follows that $\sum_{i=1}^{N}\alpha_{i}\theta_{i}(t)\equiv\theta_{p}$ and from (10b) that $\sum_{i=1}^{N}\alpha_{i}e_{i}(t)=o$. Thus we have the two relations
\begin{subequations}
\begin{numcases}
 \cdot \sum_{i=1}^{N}\alpha_{i}\theta_{i}(t)=\theta_{p}\\
 \sum_{i=1}^{N}\alpha_{i}e_{i}(t)=0
\end{numcases}
\end{subequations}
from which $\alpha_{i}$ can be computed. Further, any $m$ of the $(m+1)$ signals $e_{i}(t)$ are linearly independent.\\
{\large \underline{Comment}}: The rapidity with which $\alpha$ can be estimated from equation (11)  determines the extent to which second level adaptation will be better than first level adaptation. We note that while $\theta_{p}$ is estimated using a nonlinear equation (10b), $\alpha$ can be estimated using linear equations (11).

From equation (11) it is seen that $\alpha_{i}$ can be estimated using either equation (11a) or equation (11b). Since we treat the latter in great detail in the context of fixed models, we shall confine our attention to the first approach based on equation (11a):
\begin{equation}
\label{equation of 11a}
[\Theta(t)]\alpha=\theta_{p}
\end{equation}
where $\Theta(t)$ is the matrix whose columns are the estimates of the parameter vectors at any instant $t$ as given by the $N$ adaptive models. $\Theta(t)$ is an $[m\times (m+1)]$ matrix, $\alpha\in R^{m+1}$ and $\theta_{p}\in R^{m}$. The equation can also be expressed as
\begin{equation}
\label{reform of 11a}
\begin{split}
&\left[\begin{array}{c}\theta_{1}(t)-\theta_{m+1}(t),\theta_{2}(t)-\theta_{m+1}(t),,\cdots,\theta_{m}(t)-\theta_{m+1}(t)\end{array}\right]\\
&\cdot\left[\begin{array}{l}\alpha_{1}\\ \alpha_{2}\\ \vdots \\\alpha_{m}\end{array}\right]=\theta_{p}-\theta_{m+1}\\
&\Phi(t)\bar{\alpha}=\bar{l}(t)
\end{split}
\end{equation}
where $\Phi(t)\in R^{m\times m}, \bar{\alpha}\in R^{m}$ and $\bar{l}(t)=\theta_{p}-\theta_{m+1}(t)\in R^{m}$. $\theta_{i}(t)-\theta_{m+1}(t)$ are linearly independent for all $t$ and $i\in\{1,2,\cdots,m\}$ and in equation (\ref{reform of 11a}), both $\alpha$ and $\theta_{p}$ are unknown. However, if $\Phi(t)$ is known at two instants of time $t_{1}$ and $t_{2}$, we can solve $\bar{\alpha}$ and hence $\alpha$ in equation (\ref{adaptive case calculte alpha algebraic}). This, in turn, can be used to compute $\theta_{p}$ and the feedback control vector
\begin{equation}
\label{adaptive case calculte alpha algebraic}
[\Phi(t_{1})-\Phi(t_{2})]\bar{\alpha}=\bar{l}(t_{1})-\bar{l}(t_{2})=[\theta_{m+1}(t_{1})-\theta_{m+1}(t_{2})]
\end{equation}
An alternative approach would be to consider the derivative of $\Phi(t)$ which yields
\begin{equation}
\begin{split}
\label{}
&\dot{\Phi}(t)\bar{\alpha}=-\dot{\theta}_{m+1}(t)\\
\text{or} \ \ \ \ \  &-b^{T}P[e_{1(2)},e_{2}(t),\cdots,e_{M}(t)]\bar{\alpha}=-b^{T}Pe_{m+1}(t)\\
\text{or} \ \ \ \ \  &E(t)\bar{\alpha}=-e_{m+1}(t)
\end{split}
\end{equation}
which is the same as that obtained in the following section.

{\large \underline{Multiple Models (Fixed)}}: The principal results to come out of the previous analysis, which are contained in equation (11), are that $\sum_{i=1}^{N}\alpha_{i}\theta_{i}(t)=\theta_{p}$ at every instant of time $t$, and that $\sum_{i=1}^{N}\alpha_{i}e_{i}(t)=0$. A brief examination reveals that the latter result is also true even if the models are fixed ($i.e$ $\theta_{i}$ are constant and not estimated online). Hence once again, $\alpha_{i}$ can be determined by observing the output error vector $e_{i}(t)$ of the $N$ models.

If $\alpha_{i}$ are known, the unknown parameter vector $\theta_{p}$ can be computed using equation (11), and in turn used to compute the control parameter vector.\\
{\large \underline{Note}}: When multiple fixed models are used, the vectors $\theta_{i}$ are constant. Hence equation (11a) cannot be used to compute the vector $\alpha$, and only equation (11b) can be used.

\section{Stability analysis}
{\large \underline{Estimation of $\alpha$}}: From the proceeding discussion, it is clear that the speed of adaptation is directly proportional to the speed with which $\alpha$ can be estimated. As in conventional adaptive control, the estimate $\hat{\alpha}$ of $\alpha$ can be obtained either algebraically or using a differential equation. In both cases, the starting point is the equation $\sum_{i=1}^{m+1}\alpha_{1}e_{i}(t)=0$. This can be represented as a matrix equation as shown below:
\begin{equation}
\label{convex combination of errors}
\left[\begin{array}{c}e_{1}(t),e_{2}(t),\cdots,e_{m+1}(t)\end{array}\right]\left[\begin{array}{l}\alpha_{1}\\ \alpha_{2}\\\vdots \\\alpha_{m+1}\end{array}\right]=0
\end{equation}
with $0\leq \alpha_{i}\leq1,\sum_{i=1}^{m+1}\alpha_{i}=1$. Since $\alpha_{m+1}=1-\sum_{i=1}^{m}\alpha_{i}$, we have the matrix equation
\begin{equation}
\begin{split}
\label{matrix form of combination}
&\left[\begin{array}{c}e_{1}(t)-e_{m+1}(t),\cdots,e_{m}(t)-e_{m+1}(t)\end{array}\right]\\
&\cdot\left[\begin{array}{l}\alpha_{1}\\ \alpha_{2}\\ \vdots \\ \alpha_{m}\end{array}\right]=-e_{m+1}(t)\\
&E(t)\bar{\alpha}=-e_{m+1}(t)
\end{split}
\end{equation}

{\large \underline{Properties of the Matrix E(t)}}: The identification errors $e_{i}(t)$ of the $m$ models $(i=1,2,\cdots,m)$ were defined as
\begin{equation}
\label{definition of identification error}
e_{i}(t)=x_{i}(t)-x_{p}(t)
\end{equation}
Since all the columns of the matrix $E(t)$ are of the form $e_{i}(t)-e_{m+1}(t)$, it follows that the $i^{th}$ column of $E(t)$ is merely $x_{i}(t)-x_{m+1}(t)$ ($i.e$ the difference in the outputs of two fixed models). Hence, for any input to the $m+1$ models, the entire matrix $E(t)$ is known, and is independent of the actual plant output $x_{p}(t)$. It is only the right-hand side of equation (\ref{error for model m+1}), given by $-e_{m+1}(t)$ that depends explicitly on the plant output since
\begin{equation}
\label{error for model m+1}
-e_{m+1}(t)=x_{p}(t)-x_{m+1}(t)
\end{equation}

{\large \underline{Matrix Inversion}}: Since $\theta_{i}-\theta_{m+1}$ are linearly independent vectors for $i=1,2,\cdots,m$. It follows that the columns of $E(t)$ are linearly independent for any finite time $t$. Hence $E^{-1}(t)$ exists at every instant, so that theoretically $\alpha$ can be estimated as $\alpha=-E^{-1}(t)e_{m+1}$ in an arbitrarily short interval of time. If the plant is unstable, control action with the estimated value of $\alpha$ can be initiated at the same time instant.

\begin{figure}[H]
\includegraphics[width=0.5\textwidth,height=0.3\textwidth]{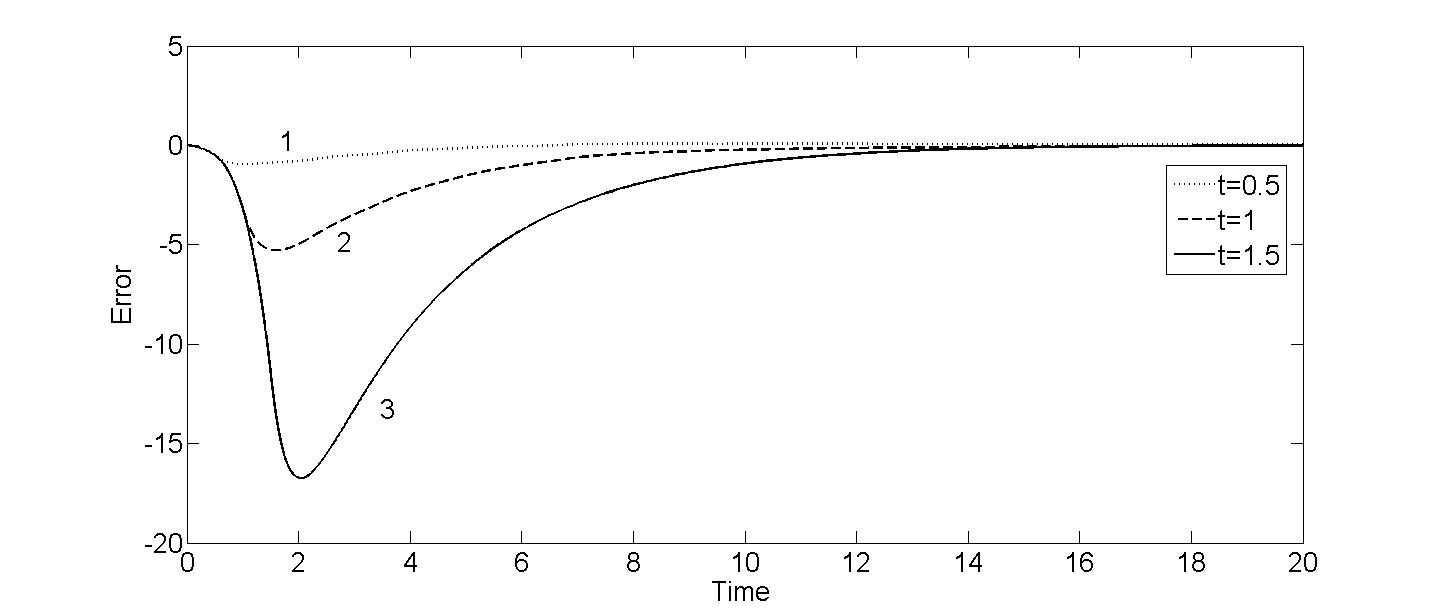}
\caption{}
\label{Fig1}
\end{figure}

\begin{Simulation}

 In view of the importance of this concept in second level adaptation, we show the errors $e_{1}(t),e_{2}(t)$ and $e_{3}(t)$ of three fixed models of an unstable second order system (with no control) over a finite period of time $T$ and the corresponding value of $\alpha_{i},(i=1,2,3)$ such that $\sum_{i=1}^{3}\alpha_{i}e_{i}(t)\equiv0,t\in[0,T]$. A stable reference model and an unstable plant are described by the parameter vectors $\theta_{m}=[-1,-3]^{T}$ and $\theta_{p}=[2,1]$. We estimate $E^{-1}(t)$ as
\begin{equation}
\begin{split}
&\left[\begin{array}{ll} -71.5919 & 5.6376 \\ 132.9577 & -10.8885 \end{array}\right],\left[\begin{array}{ll} -118.0.91 & 9.7363 \\ 199.8165 & -16.6202 \end{array}\right],\\
 \ \text{and}\ \ &\left[\begin{array}{ll} -334.5461 & 27.8656 \\ 5553.3886 &  -46.1118 \end{array}\right]
 \end{split}
\end{equation}
at three instants of time $t=0.5,1,1.5$ and compute the estimate $\hat{\alpha}(0.5)=[0.3471,0.2526,0.4003]^{T}$, $\hat{\alpha}(1)=[0.3474,0.252,0.4006]^{T}$, $\hat{\alpha}(1.5)=[0.3492,0.2491,0.4017]^{T}$.Using these values, the feedback parameters can be computed. The errors in the first state variable between plant and reference model, when estimation is carried out at $t=0.5,1,1.5$ and used to control the plant at that instant, are shown in Fig. (\ref{Fig1}).
\end{Simulation}

{\large \underline{Note}}: The above exercise is merely to indicate that the procedure is practically feasible. In practice, the estimates used are derived continuously from differential equations as shown in what follows.

{\large \underline{Estimation Using a Differential Equation}}: For well known reasons related to robustness, $\alpha$ is estimated using a differential equation rather than algebraically. In such a case $\hat{\alpha}(t)$, the estimate of $\alpha$ is determined by the differential equation
\begin{equation}
\label{differential equation for alpha}
\dot{\hat{\alpha}}=-E^{T}(t)E(t)\hat{\alpha}-E^{T}(t)e_{m+1}(t)
\end{equation}
Since the constant vector $\alpha$ satisfies the algebraic equation
\begin{equation}
E^{T}(t)E(t)\alpha+E^{T}(t)e_{m+1}(t)=0
\end{equation}
it follows that the error in the estimate $\tilde{\alpha}=\hat{\alpha}-\alpha$ satisfies the differential equation
\begin{equation}
\label{final differential equation for alpha}
\dot{\tilde{\alpha}}=-E^{T}(t)E(t)\tilde{\alpha}
\end{equation}
Using the Lyapunov function $V(\tilde{\alpha})=\frac{1}{2}\tilde{\alpha}^{T}\tilde{\alpha}$, it follows that
\begin{equation}
\frac{dV(\tilde{\alpha})}{dt}=-\|E(t)\tilde{\alpha}\|^{2}< 0
\end{equation}
or equation (\ref{final differential equation for alpha}) is asymptotically stable and $\tilde{\alpha}\to 0$. The rate of convergence depends upon the positive definite matrix $E^{T}(t)E(t)$.

When $\alpha$ is known, the plant together with the feedback gain (control) vector $\theta_{m}^{T}-\sum_{i=1}^{m}\alpha_{i}\theta_{i}$ is asymptotically stable. With $\hat{\alpha}$ in the feedback path, the overall system is represented by the differential equation
\begin{equation}
\label{plant feedback control}
\dot{x}_{p}=A_{m}x_{p}+[\Theta\tilde{\alpha}]^{T}x_{p}+br
\end{equation}
where $\Theta$ is a time-invariant matrix and $lim_{t\to\infty}\tilde{\alpha}(t)=0$. Hence, $x_{p}(t)$ tracks $x_{m}(t)$ asymptotically with zero error.
{\large \underline{Note}}: The convergence of $\tilde{\alpha}$ to zero depends only on $E^{T}(t)E(t)$ and hence on the location of the $m$ models represented by $\theta_{1},\theta_{2},\cdots,\theta_{m}$. This provides a rationale for choosing the location of the models.

{\large \underline{Comparison of First and Second Level Adaptation}}: The discussion in the proceeding section brings into better focus the difference between first and second level adaptation. In first level (or conventional) adaptive control, the emphasis is primarily on stability of the overall system. The control error is described by the equation
\begin{equation}
\label{control error differential equation}
\dot{e}_{c}=A_{m}e_{c}+b\tilde{\theta}^{T}x_{p}
\end{equation}
In attempting to stabilize the system using an approach based on the existence of a Lyapunov function, the adaptive law
\begin{equation}
\label{adaptive parameter update}
\dot{\tilde{\theta}}=-e^{T}Pbx_{p}
\end{equation}
is arrived at which is nonlinear. As is well known, the analysis of the system becomes complicated when the errors are large. In particular, when the initial conditions $\theta(t_{0})$ and $e_{c}(t_{0})$ are large, the solution of the equations (\ref{control error differential equation}) and (\ref{adaptive parameter update}) is far from simple. The only fact that has been exploited extensively is that $e_{c}^{T}Pe_{c}+\tilde{\theta}^{T}\tilde{\theta}$ is a Lyapunov function so that $e_{c}$ and $\hat{\theta}$ are bounded and $lim_{t\to\infty}e_{c}(t)=0$. However, very little can be concluded about the transient behavior of the system. Hence, additional methods, as well as arguments, are needed to justify how satisfactory response can be achieved.

 In the new approach, based on multiple models, and second level adaptation, scores significantly in such a case, since the differential equation describing the behavior of $\tilde{\alpha}$ continues to be linear. Further, the designer also has considerable prior knowledge of the matrix $E(t)$, which depends only upon the identification models which are known, rather than the plant which is unknown. This in turn, provides much greater control over the performance of second level adaptation.

 \begin{Example}
 To compare the performance of first level (conventional) adaptive control and second level adaptive control, simulation studies were carried out in the following example
\begin{equation*}
\theta_{m}^{T}=[-24,-8];\ \ \ \ \ \ \ \ \theta_{p}^{T}=[5,3]
\end{equation*}
and it is known that $\theta_{p}$ lies in the convex hull of the three models
\begin{equation*}
\theta_{1}^{T}=[-10,-10];\ \ \ \ \ \theta_{2}^{T}=[15,-10];\ \ \ \ \ \theta_{3}^{T}=[5,15]
\end{equation*}
Due to space limitations, the comparison is made only when the uncertainty is large.
 \end{Example}

\begin{figure}[H]
\includegraphics[width=0.5\textwidth,height=0.3\textwidth]{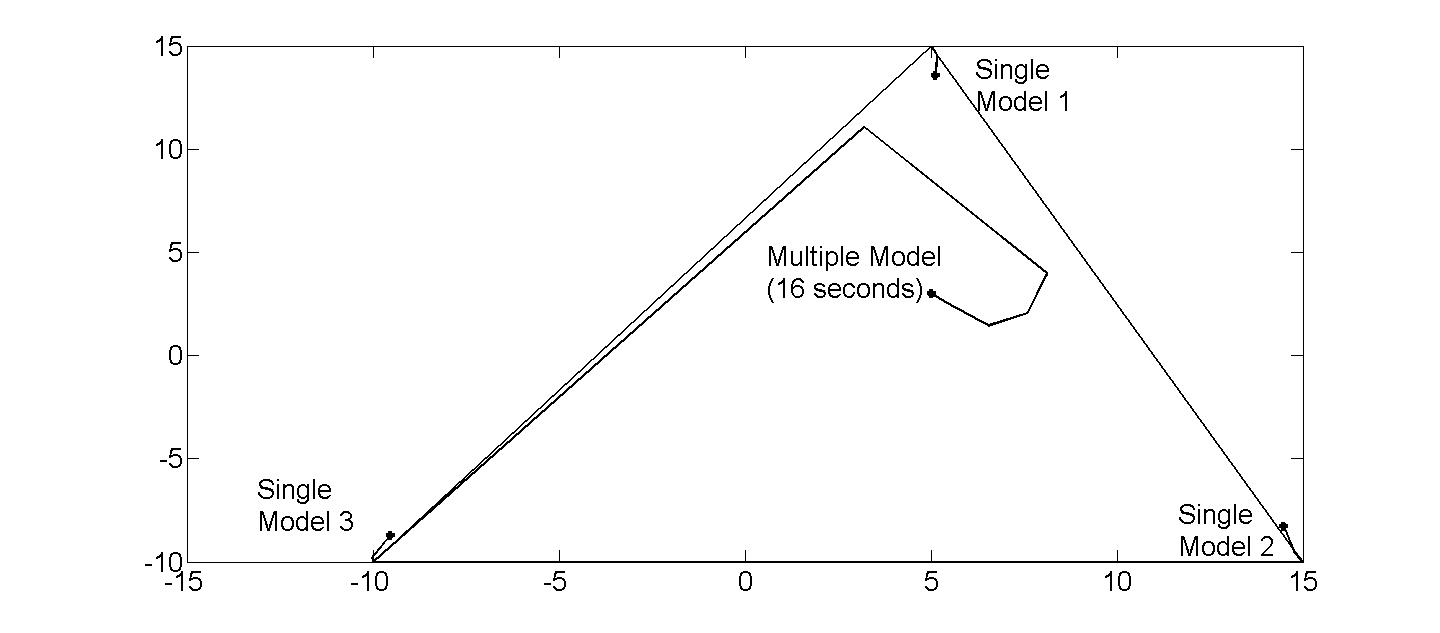}
\caption{Parameter Convergence}
\label{Fig2}
\end{figure}
\begin{figure}[H]
\includegraphics[width=0.5\textwidth,height=0.3\textwidth]{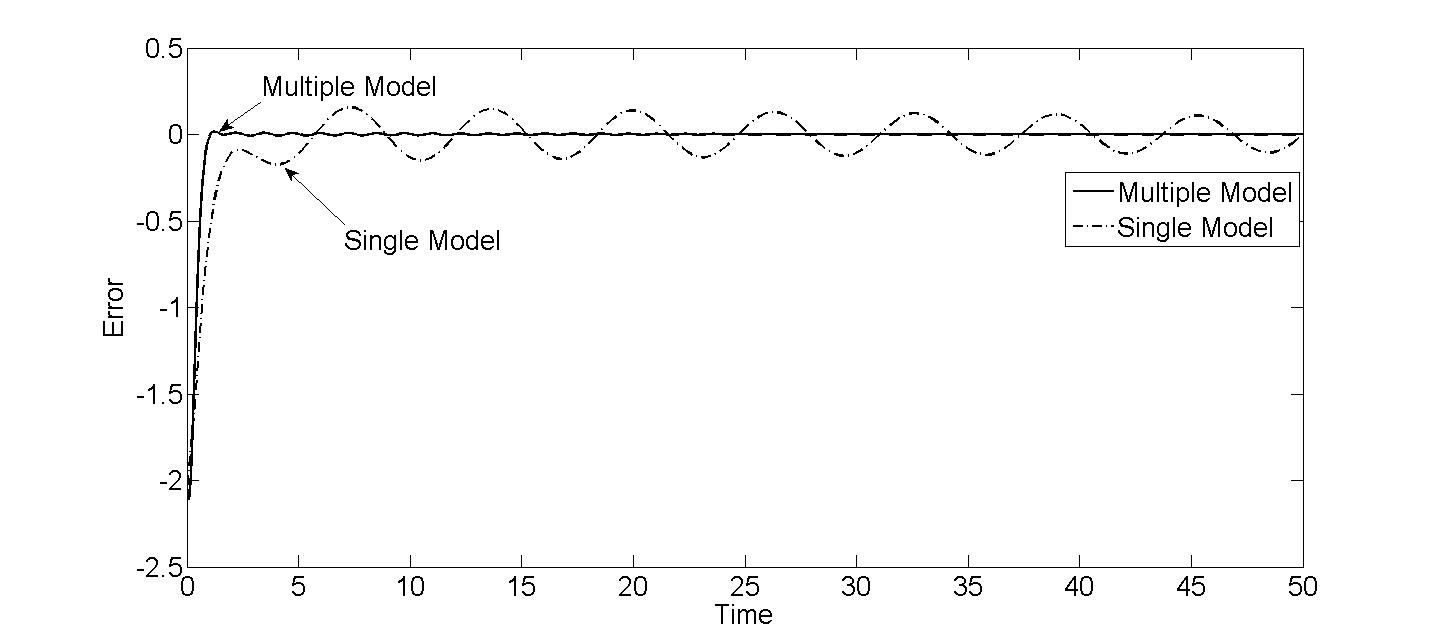}
\caption{Tracking Error}
\label{Fig3}
\end{figure}
{\large \underline{Comment}}: Fig. (\ref{Fig2}) demonstrates that the speed of convergence of second level adaptation is orders of magnitude faster than first level adaptation. While the former converges in 16 seconds, it is seen that adaptation of three first level models has barely commenced. In Fig. (\ref{Fig3}), it is seen that the tracking error (using multiple models) converges in 3 seconds, but the convergence time for the first level is considerably longer.

{\large \underline{Time Varying Parameters}}: Thus far, the analysis has been that of the adaptive control of a time-invariant plant, where $\theta_{p}$, the plant parameter vector is unknown but constant. We have discussed the advantage of using second level adaptation over conventional adaptive control when the region of uncertainty is large. In such a case, the  nonlinear effects in the adaptive algorithm (first level) begin to play a major role in the convergence of the parameter estimate. In this section, we deal with time-varying plant parameters, which are becoming increasingly important in adaptive control.

Let $\theta_{p}(t)$ be a time-varying plant parameter vector, which lies in the convex hull of the fixed models $\theta_{1},\theta_{2},\cdots,\theta_{m+1}$. Since $\sum_{i=1}^{m+1}\alpha_{i}\theta_{i}=\theta_{p}(t)$, it follows that $\alpha(t)$ must be time-varying. Since $\theta_{i},(i=1,2,\cdots,m+1)$ are constant,
\begin{equation}
\label{parameter combination for time varying system}
\begin{split}
&\left[\begin{array}{c}\theta_{1}-\theta_{m+1},\theta_{2}-\theta_{m+1},\cdots,\theta_{m}-\theta_{m+1}\end{array}\right] \bar{\alpha}(t)\\ &=\theta_{p}(t)-\theta_{m+1}\\
\text{or} \ \ &\Theta\bar{\alpha}(t)=\theta_{p}(t)-\theta_{m+1}
\end{split}
\end{equation}
While the equation (\ref{parameter combination for time varying system}) cannot be used to determine $\alpha$, it nevertheless indicates that $\alpha(t)$ can be estimated at least as rapidly as the variations  in $\theta_{P}(t)$.

{\large \underline{First Level Adaptation}}: The behavior of adaptive systems with time-varying parameters has been the subject of the research for several decades. However, useful results are available only when $\theta_{p}$ has rapid variation with small amplitude or slow but large variations. Since $\tilde{\theta}=\hat{\theta}(t)-\theta_{p}(t)$, $\dot{\tilde{\theta}}(t)=\dot{\hat{\theta}}(t)-\dot{\theta}_{p}(t)$. In other words, $\dot{\theta}_{p}(t)$ is a forcing function of the nonlinear error differential equation. This accounts for the difficulty encountered in analyzing adaptive systems with time-varying parameters.

{\large \underline{Second Level Adaptation}}: The simple equations (11a) and (11b), which can be used to determine $\alpha$, were discussed assuming linearity and time-invariance of the error equations. These in turn were used to generate the differential equation (\ref{differential equation for alpha}) for determining $\hat{\alpha}$ and consequently the control parameter.

In the time-varying case, the analysis is considerably more complex, but arguments can be provided as to why $\sum_{i=1}^{m+1}\alpha_{i}(t)e_{i}(t)=0$ is still a valid approximation of the equation (11b).
In the following analysis, we assume that equation $\sum_{i=1}^{m+1}\alpha_{i}(t)e_{i}(t)=0$ is a valid approximation of the equation (11a). Once again, we use $\hat{\alpha}(t)$, the solution of the differential equation
\begin{equation}
\label{alpha differential equation}
\dot{\alpha}(t)=-E^{T}(t)E(t)(t)\hat{\alpha}-E^{T}(t)e_{m+1}
\end{equation}
to approximate the time varying $\alpha(t)$. Since $E(t)$ depends only on the output of the fixed identification models, it is not affected by the time variations in the plant parameters. This is only reflected in the term $E^{T}e_{m+1}(t)$ where $e_{m+1}(t)$ is the output error between the $(m+1)_{th}$ model and the plant. Since $-E^{T}(t)E(t)$ is an asymptotically stable matrix, the effect of  parameters variations are reflected as bounded variations in $\hat{\alpha}(t)$, the estimate of $\alpha(t)$. From the qualitative description of the time-varying problem given above, it is evident that the analysis of the effect of time-varying parameters is considerably simpler than in conventional adaptive control. As in all linear systems, the convergence of the adaptive algorithm can be adjusted using a simple gain parameter.
\begin{Example}
Using both first and second level adaptation, the following adaptive control problems with time-varying parameters were simulated. The plant is of third order with parameter $\theta_{p}(t)$. Two specific cases were considered as described below:

\begin{equation*}
\begin{split}
\text{Experiment 1}:\ \ \ \theta_{p}&=[3+sin(0.5t),4+cos(0.5t),3]\\
 \text{Experiment 2}:\ \ \ \theta_{p}&=[3+f(t),4+f(t),3]
\end{split}
\end{equation*}
where $f(t)$ is a square wave with mean value zero, amplitude 1 and period of 40 units.

In Experiment 1 the time-varying parameters vary sinusoidally, while in Experiment 2 they vary discontinuously over intervals of 10 units. In both cases, the objective is to track a reference signal, which is the output of a reference model with parameter $\theta_{m}^{T}=[-15,-23,-9]$.

The variation of $\theta_{p}(t)$ in $R^{2}$ is shown Fig. (4a); the variations of $f(t)$ as a function of time in Experiment 2 is shown in Fig. (4b).

The response due to first level adaptation and second level adaptation in Experiment 1 are shown in Fig. (5a) $\&$ (5b) and that for Experiment 2 are shown in Fig. (6a) $\&$ (6b). Fig. (7) show the tracking of the true value of parameter $\theta_{p}$ in Experiment 2 using first level adaptation and second level adaptation. In all cases, second level adaptation is seen to result in significantly better response that first level adaptation.

\begin{figure}[H]
  \centering
  \includegraphics[width=0.5\textwidth ]{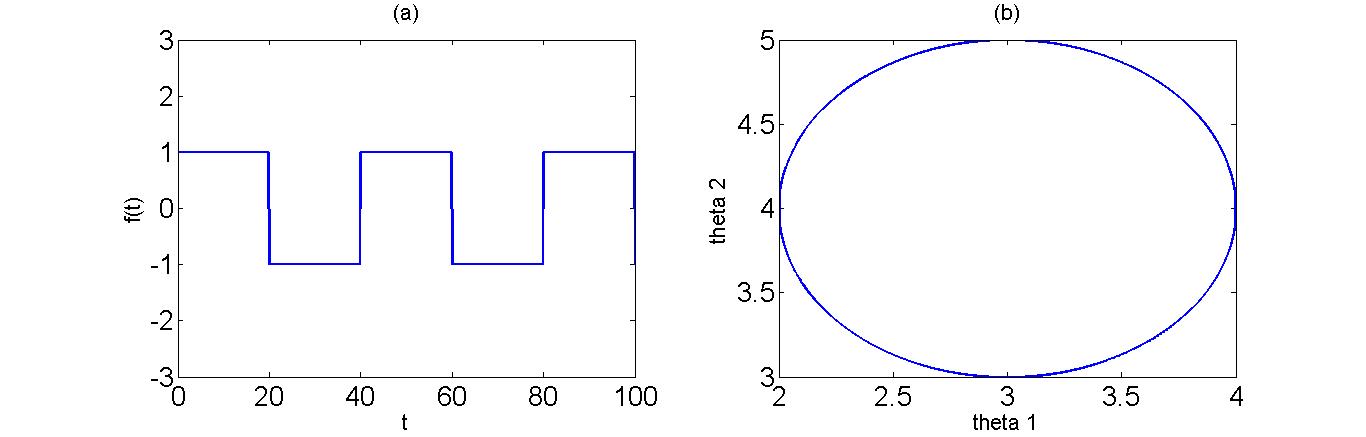}\\	
  \caption{The parameter variation}
  \label{Fig4}
\end{figure}

\begin{figure}[H]
\centering
\subfigure[Single Model]{
\begin{minipage}[b]{0.5\textwidth}
\includegraphics[width=1\textwidth]{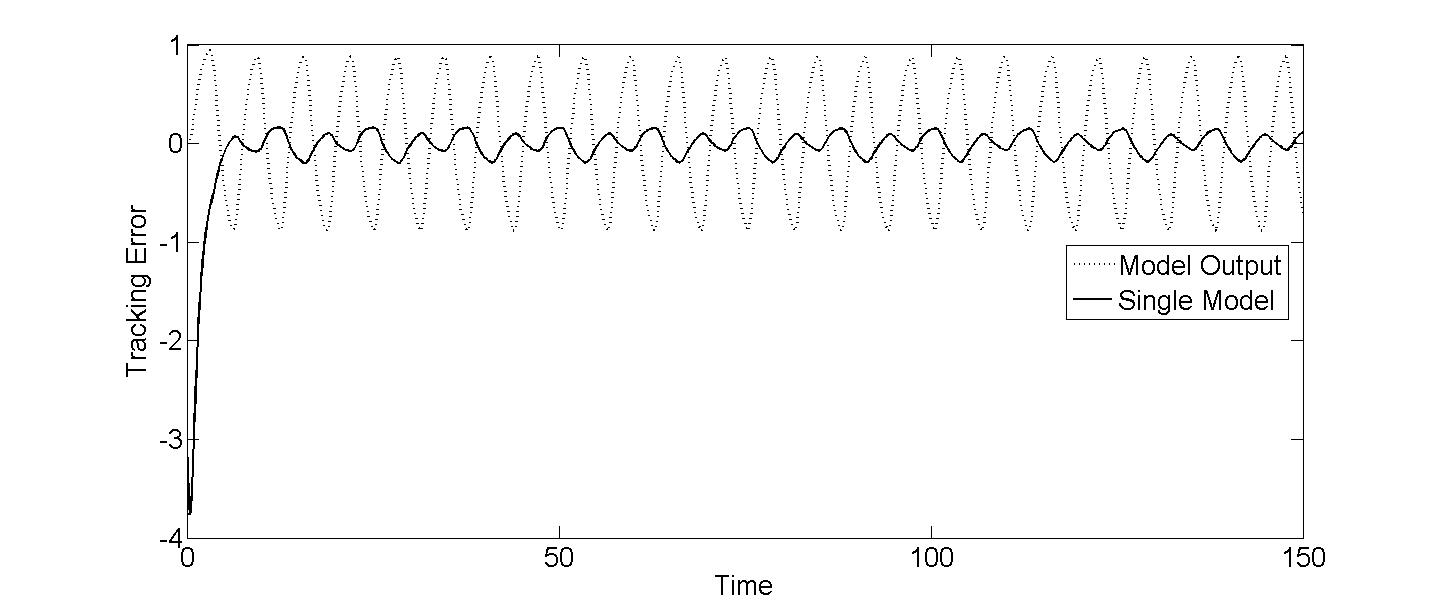} \\
\end{minipage}
}
\subfigure[Multiple Model]{
\begin{minipage}[b]{0.5\textwidth}
\includegraphics[width=1\textwidth]{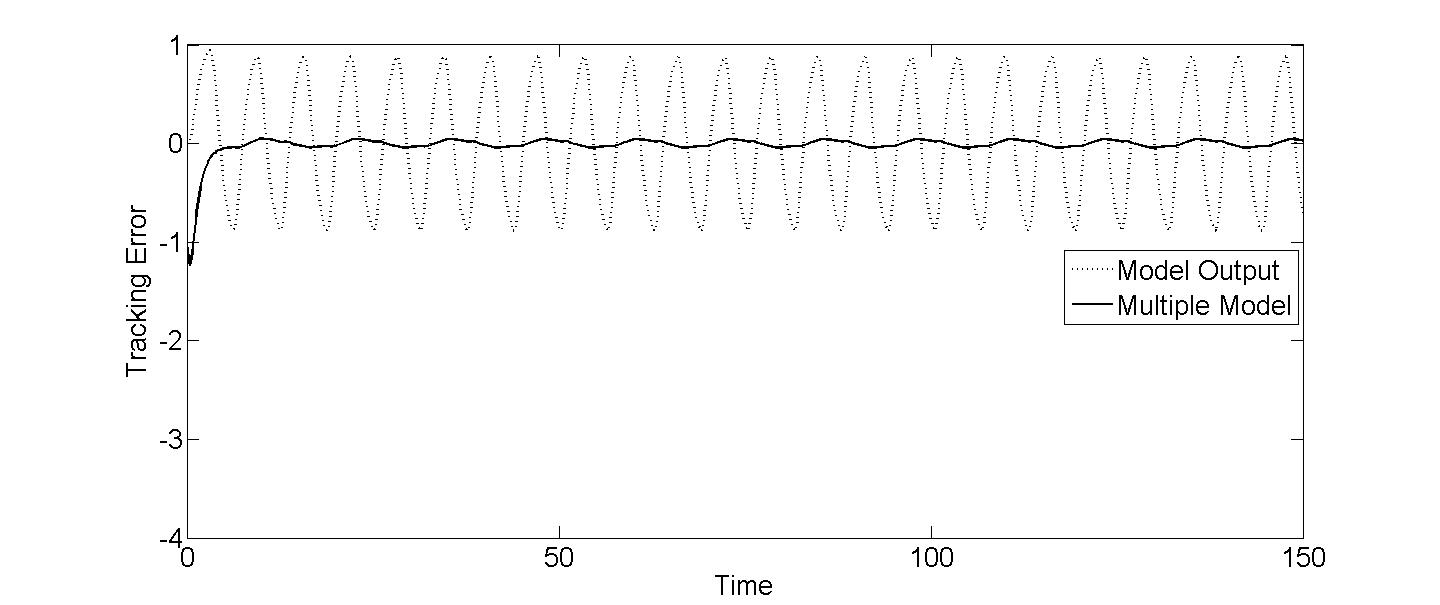}
\end{minipage}
}
\caption{Output Tracking Error of Experiment 1}
\end{figure}

\begin{figure}[H]
\centering
\subfigure[Single Model]{
\begin{minipage}[b]{0.5\textwidth}
\includegraphics[width=1\textwidth]{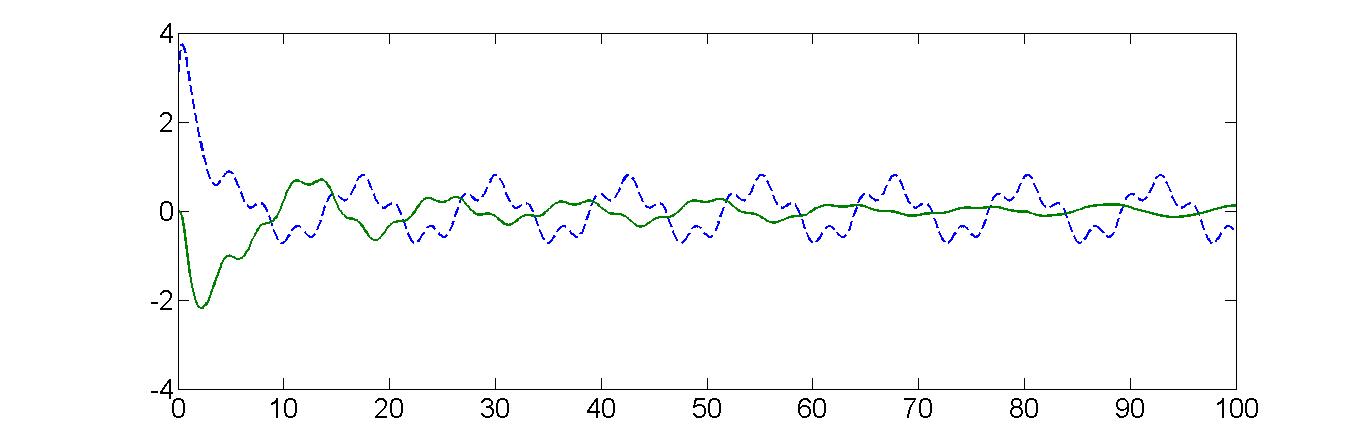} \\
\end{minipage}
}
\subfigure[Multiple Model]{
\begin{minipage}[b]{0.5\textwidth}
\includegraphics[width=1\textwidth]{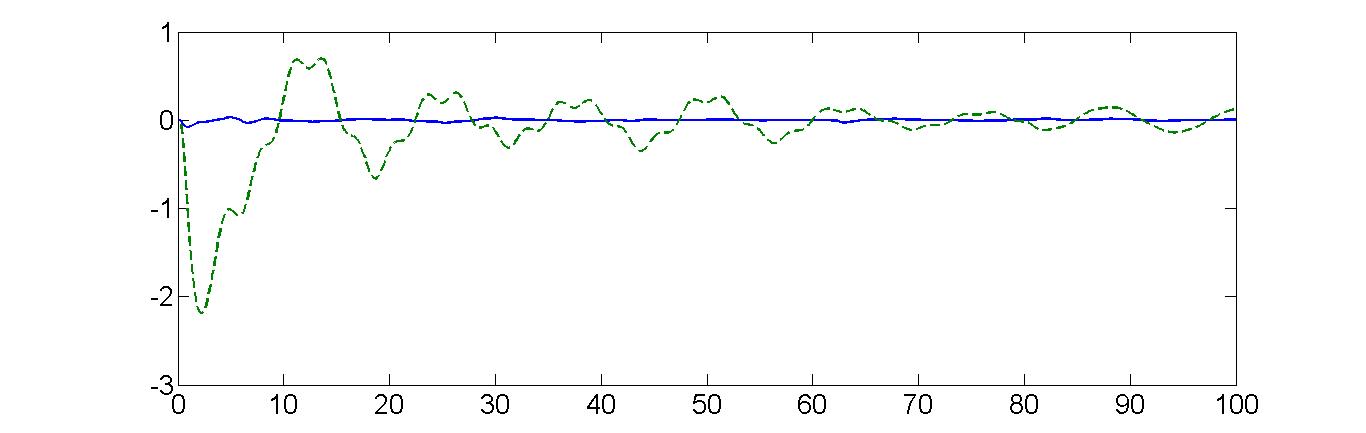}
\end{minipage}
}
\caption{Output Tracking Error of Experiment 2}
\end{figure}

\begin{figure}[H]
\centering
\subfigure[Single Model]{
\begin{minipage}[b]{0.5\textwidth}
\includegraphics[width=1\textwidth]{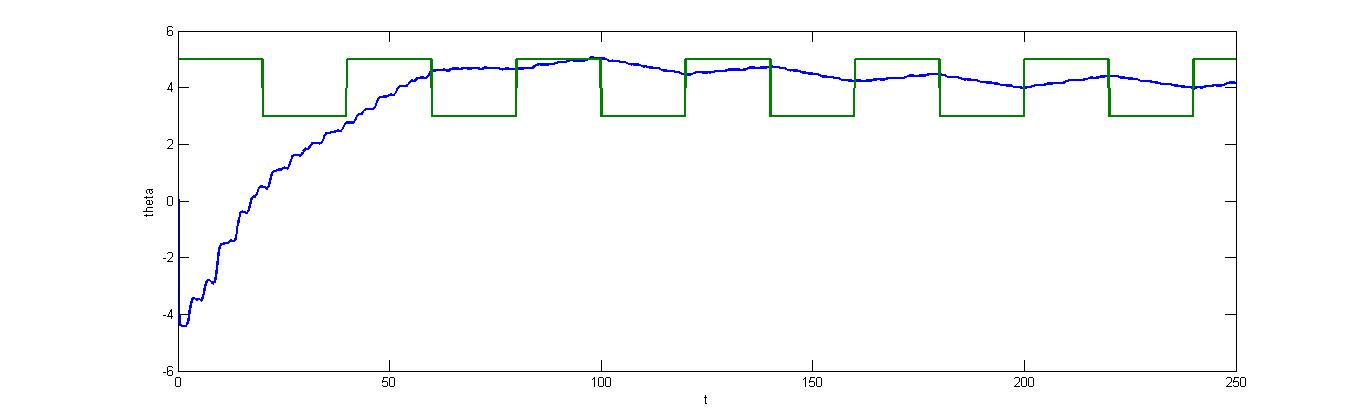} \\
\end{minipage}
}
\subfigure[Multiple Model]{
\begin{minipage}[b]{0.5\textwidth}
\includegraphics[width=1\textwidth]{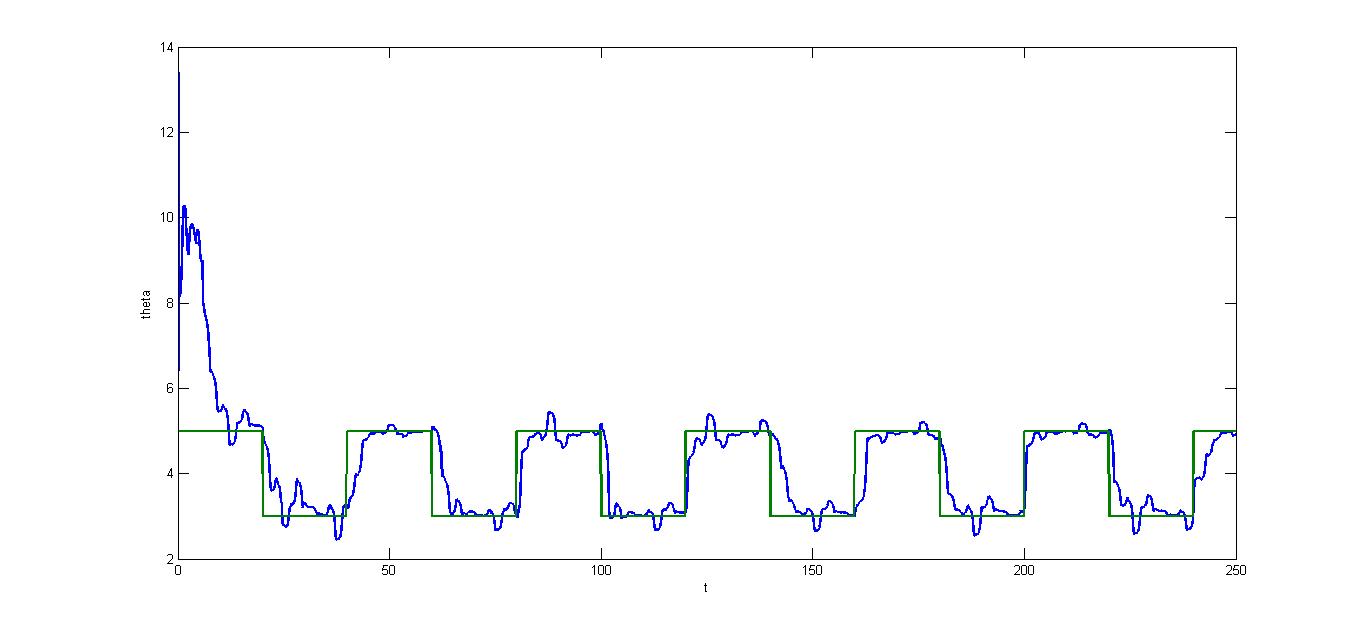}
\end{minipage}
}
\caption{Parameter Tracking Error of Experiment 2}
\end{figure}

\end{Example}

\section{Fixed Model and Adaptive Models}
It was seen from the discussions in earlier sections that either fixed models or adaptive models can be used for identifying and controlling a plant. Extensive simulation studies have shown that multiple adaptive models (rather than multiple fixed models) result in a smoother response. However, the convergence times are comparable in the two cases. If the plant parameter is not constant but time-varying, the approach using fixed models is found to be better, since in the long run, all adaptive models tend to converge to a single point in parameter space. This make the approach almost equivalent to conventional adaptive control. Hence, depending on the application either fixed or adaptive models can be used in second level adaptation.

The use of fixed and adaptive models for switching and tuning is a well investigated approach for adapting to large uncertainty. The experience gained from those studies are currently being used to determine how the two types of models can be combined for maximum effect in second level adaptation.

\section{Robustness of Second Level Adaptation}
The behavior of adaptive systems with different types of perturbations has been studied for decades under the title of "Robustness". In particular,
\begin{itemize}
\item [(\expandafter{\romannumeral1})] the effect of input output disturbance
\item [(\expandafter{\romannumeral2})] tine variation in unknown parameters
\item [(\expandafter{\romannumeral3})] the effect of unmodeled dynamics
\end{itemize}
on the performance of the adaptive system have to be studied in this context.

We have already seen that second level adaptation performs significantly better than first level, both when the region of uncertainty is large and when the plant parameter varies rapidly with time.

When the output of the plant is corrupted with noise, it is seen that the matrix $E(t)$ is unaffected while the signal $e_{m+1}(t)$ contains noise. This results in $\tilde{\alpha}(t)$ having zero mean value due to the noise and hence smaller error in the response of the adaptive system. Hence, second level adaptation is found to be more robust than first level adaptation in two out of three cases of interest. The behavior of second level adaptation in the presence of unmodeled dynamics is currently being investigated.

\section{Conclusion}
Second level adaptation based on multiple identification models appears to have advantages over conventional or first level adaptive control in regard to every feature of interest to the designer. It is significantly faster with large uncertainties, and/or rapid time-variation of parameters, and more robust in the presence of disturbances. All the advantages are due to the fact that $\alpha$, which is an alternative parameterization of $\theta_{p}$ can be estimated using linear equations while $\hat{\theta}_{p}(t)$ is invariable determined by nonlinear equations, due to the importance of stability in adaptive systems.

All the analysis carried out in the paper deals with the plant in companion form, whose state variables are accessible for measurement. Extensions to more general representations, and adaptation using only the input-output of the plant are currently  being investigated.

\end{document}